# Local intermolecular structure, hydrogen bonding and related dynamics in the liquid cis/trans N-methylformamide mixture: A density functional theory based Born-Oppenheimer molecular dynamics study.

Ioannis Skarmoutsos[1,*], Ricardo L. Mancera[2], Stefano Mossa[3] and Jannis Samios[4]

[1] *Laboratory of Physical Chemistry, Department of Chemistry, University of Ioannina, 45110 Ioannina, Greece*
[2] *Curtin Medical School, Curtin Health Innovation Research Institute and Curtin Institute for Computation, Curtin University, GPO Box U1987, Perth, Western Australia 6845, Australia*
[3] *CEA, IRIG-MEM, Univ. Grenoble Alpes, 38000 Grenoble, France*
[4] *Department of Chemistry, Laboratory of Physical Chemistry, National and Kapodistrian University of Athens, Panepistimiopolis 157-71, Athens, Greece*

## Abstract

The local intermolecular structure and related dynamics in the liquid cis/trans N-methylformamide mixture at ambient temperature and density conditions have been systematically studied by employing Born-Oppenheimer molecular dynamics simulation techniques. Particular attention has been paid to the local structure around the cis- and trans- conformers and the formation and dynamics of hydrogen bonds with their closest neighbors. The calculated atom-atom radial distribution functions are in very good agreement with available experimental data and reveal the existence of different types of hydrogen bonding intermolecular interactions. The average number of hydrogen bonds formed by the cis- conformers is higher in comparison with the one corresponding to the trans- conformers. Moreover, the lifetimes of the hydrogen bonds formed in the liquid are longer when the cis- conformers participate in the bond formation, either as donors or acceptors. These findings clearly indicate that the local structural network around the cis- conformers in the liquid is more cohesive. The latter finding is also reflected in the slower reorientational dynamics of the cis- conformers and the low- and high-frequency region of the spectral densities of the atomic velocity time correlation functions. Finally, the calculated average dipole moments of the trans- and cis- conformers are significantly higher than their corresponding gas-phase values, signifying the importance of polarization effects in this particular polar liquid solvent.

[*] Corresponding author: iskarmoutsos@uoi.gr

**Graphical Abstract**

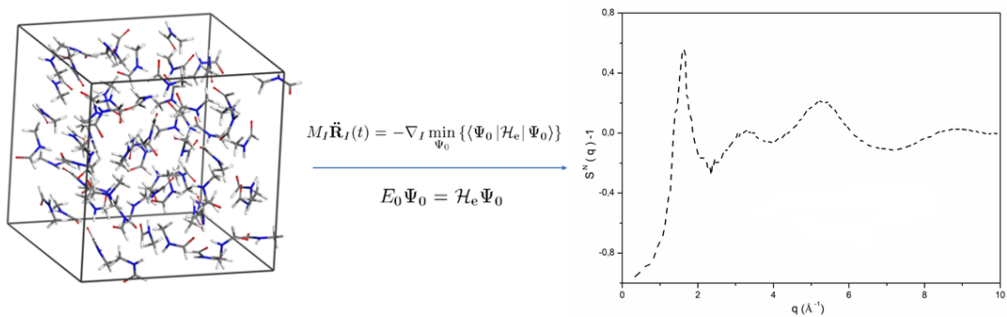

**Keywords:** Liquid N-methylformamide, Conformers, Born-Oppenheimer MD, Liquid structure, Hydrogen bonding, Reorientational dynamics

**Highlights**

- The cis/trans NMF liquid mixture has been studied using Born-Oppenheimer MD simulations.
- The calculated RDF and neutron-weighted structure factor are in good agreement with experimental data.
- The average number of hydrogen bonds formed by the cis- conformers is higher in comparison with the one corresponding to the trans- conformers.
- The local HB network around the cis- conformers is maintained for longer timescales, causing a slowing down of their reorientational dynamics.
- The calculated average dipole moments of the trans- and cis- conformers are significantly higher than their corresponding gas-phase values.

# 1. Introduction

Liquid amides represent a very important class of solvents due to their use in several applications in chemical science and technology [1,3]. From a molecular point of view, the simplest amide molecules can also be considered as prototypes of peptides and proteins, since the hydrogen bonding (HB) ability of the amide group is very important in determining their secondary and tertiary structure [4-6]. Molecules containing the amide bond have a wide range of applications in the pharmaceutical industry and the molecular-scale understanding of their interactions with biological targets is a key factor in molecular biology and drug discovery [7]. The high polarity of protic amide solvents, as well as their ability to form hydrogen bonds, makes them especially suited for investigating the cooperative dynamics of hydrogen-bonded, associated liquid systems using a wide range of experimental and theoretical-simulation techniques [8-11].

For all these reasons, the simplest hydrogen-bonded liquid amides, such as formamide (FA), N-methylformamide (NMF) and N-methylacetamide (NMA) have been investigated in the past by experimentalists and theoreticians, aiming to provide a more detailed description of the local structure and dynamics in these liquid solvents [8-17]. Particularly liquid NMF, which exhibits a very high dielectric constant [18], has been the topic of several studies [19-29]. Interestingly, according to several experimental observations, liquid NMF consists of cis- and trans- conformers [30-37]. The mole fraction of the cis-conformers in the liquid mixture, $X_{cis}$, has been estimated to be in the range 0.06-0.1 [30,35]. In the earliest attempt to simulate liquid NMF as a two-component mixture, Chen et al. employed a polarizable force field in Monte Carlo simulations to model the cis- and trans- conformers [30]. However, the model employed in that study exhibited some deviations between the calculated thermodynamic properties of the mixture and experimental values. For this reason, in a subsequent molecular dynamics (MD) treatment [3] we developed and proposed a non-polarizable, all-atom potential model for both conformers, including only Lennard-Jones and Coulombic interactions. Using that potential model, we predicted accurately the thermodynamic properties of the mixture and provided for the first time in the literature some insight on the effect of the local intermolecular structure around the two conformers on their dynamics. The translational dynamics of the trans-NMF molecules was found to be slower than the corresponding dynamics of the cis-NMF conformers, in agreement with experimental data [30]. We also revealed, using information obtained by the calculated radial distribution functions (RDFs), that the existence of HB interactions between the oxygen and carbonyl hydrogen atoms is less favored in comparison with the $H_N\ldots O$ hydrogen bond. However, we pointed out that this weak correlation between the oxygen and carbonyl-hydrogen atoms leads to the stabilization of such kind of HB dimer configurations.

A few years after the publication of our work, neutron diffraction studies combined with empirical potential structure refinement (EPSR) simulations [25] provided novel insight on the intermolecular structure of liquid NMF. In that work, the atom-atom RDFs corresponding to the experimental structure factor F(q) were presented, revealing more pronounced HB interactions in comparison with our previous classical MD simulation studies. The more pronounced HB interactions were reflected in the peak positions and intensities of the obtained $H_N$ - O and $H_C$ - O RDFs. According to the EPSR simulations, dimers and trimers of hydrogen-bonded molecules form clusters of up to 5 Å in radius. The existence of weaker $H_C \ldots O$ hydrogen bonds was also revealed in the EPSR simulations. The existence of this kind of HBs, according to the authors of that study, may play an important role in the stabilization of the intermolecular structure in liquid NMF, a statement that is in agreement with our previous findings. However, in the EPSR simulations only the trans-conformers were taken into acount and the local structure around the cis-conformers was not investigated.

Taking all the above into account, the main aim of the present study is to further investigate the local HB structure and corresponding dynamics in the liquid cis/trans NMF mixture, using density functional theory (DFT) based *Born-Oppenheimer* molecular dynamics [38] (BOMD) simulations. HB is a complex type of interaction [39], with contributions arising from electrostatic, induction and dispersion interactions, exchange correlation effects from short-range repulsion due to overlap of the electron distribution and charge-transfer-induced covalency. Therefore, a detailed description of these interactions in a quantum-mechanical framework, as attempted in the present study via BOMD simulations, is required in order to deeper investigate HB interactions and related dynamics in liquid solvents.

## 2. Computational Methods

BOMD simulations of the liquid cis/trans NMF mixture at 298.15 K and a density of 0.9985 $g/cm^3$ were performed using a cubic simulation box containing 58 trans-NMF and 6 cis-

NMF molecules, respectively. The simulated mixture was initially equilibrated by performing a classical MD simulation, using the DL_POLY code [40], in the isothermal-isobaric (NPT) ensemble with P=1 bar for a period of 5 ns, using the potential models developed for trans- and cis-conformers in our previous study [3]. Using the calculated density of the mixture (0.9985 g/cm$^3$), which comes in excellent agreement with the experimental value [3], a subsequent classical MD simulation in the canonical ensemble (NVT) was performed for a period of 10 ns in order to further equilibrate the system and prepare the starting configuration for the BOMD simulations. A NVT-BOMD equilibration run was subsequently performed for a period of 20 ps, followed by a production period of 80 ps for the calculation of the properties of the mixture. It should be also noted that during the BOMD simulation, intramolecular rotations around the central C-N axis, leading to transitions from the cis- to the trans- conformers and vice-versa, were not observed. This observation is in agreement with the previous findings of Chen et al [30], who analyzed the data obtained by previous nuclear magnetic resonance (NMR) experiments [31] and estimated the lifetime of the cis-NMF conformers to be around 60 s, which is much longer than the BOMD simulation timescale (100 ps).

The density functional employed in the present study was the PBE [41], including the D3 dispersion correction suggested by Grimme et al (PBE+D3) [42]. A hybrid Gaussian and plane waves methodology, as implemented in the Quickstep [43] part of the cp2k code [44], was employed in the BOMD simulations. The DZVP-MOLOPT basis set [45] and the Goedecker-Teter-Hutter (GTH) [46] pseudopotentials were used for the electronic structure calculations. The density cutoff was set to 480 Ry to yield well converged calculations. During the 20 ps equilibration run, the thermal equilibration was achieved using the canonical scaling via velocity rescaling (CSVR) thermostat [47] with a strong coupling. For the 80 ps NVT-BOMD production run, the canonical ensemble was sampled using a weaker coupling to the bath. The equations of motion were integrated via the Velocity-Verlet algorithm [48], using a 0.5 fs time step. We also have to note that test classical MD runs, varying the simulated system size and time-scale, have ensured that the particular size and time-scale of the BOMD simulation provides a reliable statistical sampling of the investigated properties of interest.

**3. Results and Discussion**

## 3.1 Local Intermolecular Structure – Hydrogen Bonding

The local intermolecular structure in the cis/trans liquid NMF mixture was initially investigated in terms of the center of mass (COM) pair RDFs. The calculated COM-COM RDFs for the trans-trans, cis-trans and cis-cis NMF pairs are presented in Figure 1, clearly depicting the differences in the local solvation structure around the cis- and trans-NMF conformers. The higher intensity of the first peak of the cis-cis RDF, located at about 4.8 Å, clearly indicates a preference of the cis- conformers to interact with other cis- conformers, although their concentration in the mixture is low. This is also quantitatively reflected in the radial dependence of the corresponding average local mole fraction of cis-conformers around each individual cis- conformer, $X^{cis\text{-}cis}(r)$ [49], which at distances around 4.8 Å exhibits a value close to 0.14, about 1.55 times higher than the bulk mole fraction of the cis- conformers used in the simulation. The first solvation shell around the cis- and trans- conformers is located at about 6.8 Å. The coordination number of the trans- and cis- conformers which are inside the first solvation shell of trans- conformers are 11.7 and 1.2, respectively, yielding an overall coordination number of 12.9. We note that the EPSR simulations of Cordeiro and Soper [25] estimated that the coordination number of trans- conformers inside the first solvation shell of trans- conformers, with a 6.7 Å radius, is 11.9. This finding is in very good agreement with the results obtained by the present BOMD simulations. The corresponding coordination number of the trans- and cis-conformers inside the first solvation shell of cis- conformers are 11.5 and 1.2, respectively, yielding an overall coordination number of 12.7.

In the framework of the present study the neutron-weighted static structure factor $S^N(q)$ [50-53], which can be measured by neutron diffraction experiments, was calculated using the following expression:

$$S^N(q) = \left\langle \frac{N}{\sum_\alpha N_\alpha b_\alpha^2} \sum_\alpha \sum_\beta b_\alpha b_\beta S_{\alpha\beta}(\vec{q}) \right\rangle \quad , \quad q = |\vec{q}| \tag{1}$$

where $b_\alpha$ is the coherent neutron scattering length for species α. Here, the brackets indicate a spherical average over wave vectors of modulus q, and the partial static structure factors involving species α and β are defined as

$$S_{\alpha\beta}(\vec{q}) = \frac{(1+\delta_{\alpha\beta})}{2N} \rho_\alpha(\vec{q})\rho_\beta^*(\vec{q}) ,\qquad(2)$$

$$\rho_\alpha(\vec{q}) = \sum_{l=1}^{N_\alpha} \exp(i\cdot\vec{q}\cdot\vec{r}_l) ,\qquad(3)$$

with $\vec{r}_l$ being the instantaneous vector position of atom $l$.

The calculated structure factor $S^N(q)-1$ is presented in Figure 2 and exhibits the same shape as the experimental F(q) factor presented by Cordeiro and Soper (Figure 2 of Reference 25), with the local maxima and minima also located at exactly the same wave vectors q. This finding clearly indicates that the BOMD simulation provides reliable descriptions of the liquid structure of the mixture. This is more clearly reflected upon the shape and intensities of the calculated atom-atom RDF for the trans-trans NMF pairs, presented in Figure 3, which are in very good agreement with the ones obtained by the EPSR simulations of Cordeiro and Soper (Figure 4 of Reference 25). Comparison of the shape and peak intensities of the O-$H_N$, O-$H_C$ and O-$H_{Me}$ RDFs, clearly reveals that the formation of O…$H_N$ hydrogen bonds is more pronounced than the O…$H_C$ ones. On the other hand, the O…$H_{Me}$ HB interactions are much weaker, which is clearly reflected in the appearance of a weak shoulder in the corresponding RDF at short intermolecular distances around 2.5 Å and a first minimum located at 5.2 Å.

The calculated cis-trans RDFs that are related to HB interactions are presented in Figure 4. It can be clearly observed that the formation of O…$H_N$ hydrogen bonds is the most pronounced, particularly in the case when cis- conformers molecules act as HB acceptors. The formation of O…$H_C$ hydrogen bonds also depends on whether the cis- conformers molecules act as HB donors or acceptors, something which is clearly reflected on the different shape of the O-$H_C$ and $H_C$-O cis-trans RDFs. From the shape of the calculated RDFs it can also be clearly seen that the O…$H_{Me}$ hydrogen bonds are very weak in this case as well, independently of whether the cis- conformers molecules act as HB donors or acceptors.

The cis-cis RDFs that are related to HB interactions are presented in Figure 5. Here we can clearly observe that the formation of O…$H_N$ hydrogen bonds is significantly more pronounced over all the other types of HB interactions, as clearly reflected on the high intensity of the first peak of the O-$H_N$ cis-cis RDF at 1.8 Å. The formation of O…$H_N$ hydrogen bonds between the cis- conformers is therefore the main reason for the high-intensity peak of the COM-COM cis-cis RDF and the increase of the local mole fraction of cis- conformers around each individual cis- conformer at short intermolecular distances.

To provide a more quantitative description of the formation of hydrogen bonds in the liquid cis/trans NMF mixture, we decided to perform a HB analysis based on geometric criteria using information from the shape of the calculated atom-atom RDF, as commonly employed in the literature for a wide range of hydrogen bonded fluids [54,55]. In the present study, based on the shape of the calculated atom-atom RDF, we focused on the formation of O…$H_N$ and O…$H_C$ hydrogen bonds, since we observed that the O…$H_{Me}$ HB interactions are very weak. According to the geometric criterion used to describe the formation of $H_N$…O hydrogen bonds, a hydrogen bond between two NMF molecules (cis- or trans-) exists if the interatomic distances are such that $R(N^{donor}…O^{acceptor}) \leq 3.43$ Å, $R(H_N…O) \leq 2.78$ Å and the donor-acceptor angle $H_N$-N…O $\leq 30º$ (the symbol - corresponds to intramolecular distances and the ... symbol to intermolecular ones). Similarly, using the same criterion, in the case of $H_C$…O hydrogen bonds, a hydrogen bond between two NMF molecules (cis- or trans-) exists if the interatomic distances are such that $R(C^{donor}…O^{acceptor}) \leq 3.43$ Å, $R(H_C…O) \leq 2.78$ Å and the donor-acceptor angle $H_C$-C…O $\leq 30º$.

The calculated average number of hydrogen bonds formed per trans- or cis- conformers, for each different case where the trans- or cis- conformers act as HB donors or acceptors, are presented in Tables 1,2. From Table 1 we may see that the total average number of $H_N$…O and O…$H_N$ hydrogen bonds formed by a trans- conformer (in cases where the central trans- conformer acts either as HB donor or acceptor) is 1.6. On the other hand, the total number of $H_C$…O and O…$H_C$ hydrogen bonds formed by a trans- conformer, when acting either as HB donor or acceptor, is 0.4. Therefore, we may conclude that according to our HB analysis each trans- conformer totally forms on average two hydrogen bonds with its closest trans- conformer neighbors, when acting either as a HB donor or acceptor. However, the formation of $H_N$…O and O…$H_N$ hydrogen bonds is significantly more pronounced, as also revealed by the calculated atom-atom RDF. On the other hand, each trans- conformer forms on average 0.2 hydrogen bonds with its closest cis- conformer neighbors, which are also mainly $H_N\ldots O$ and $O\ldots H_N$ hydrogen bonds (an average of 0.15 hydrogen bonds of this type in total).

From Table 2 we may also see that each cis- conformer forms on average 0.54 hydrogen bonds with its closest cis- conformer neighbors, which are mainly $H_N\ldots O$ and $O\ldots H_N$ hydrogen bonds (resulting in an average of 0.5 hydrogen bonds of this type). This is an additional quantitative indication of the importance of cis-cis HB interactions in the mixture. On the other hand, the total average number of cis-trans hydrogen bonds formed per cis-NMF molecule is 1.91. The average numbers of $H_N\ldots O$ and $O\ldots H_N$ cis-trans hydrogen bonds formed per cis- conformer are 0.65 and 0.79, respectively, signifying that this type of cis-trans HB interactions is more pronounced when the cis- conformers act as HB acceptors. Finally, the average numbers of $H_C\ldots O$ and $O\ldots H_C$ cis-trans hydrogen bonds formed per cis- conformer are 0.27 and 0.20, respectively, indicating that HB interactions among the carbonyl hydrogen atom and the oxygen atom are slightly more pronounced when the cis- conformers act as HB donors. We can generally state that the total number of hydrogen bonds formed per cis-NMF molecules with its closest cis- and trans-NMF neighbors is 2.45, whereas the corresponding total average number of hydrogen bonds per trans- conformer is 2.2. The latter prediction is also in agreement with the EPSR simulations of Cordeiro [25], who found that a trans- conformer is on average hydrogen-bonded to two other trans- conformer molecules. All these findings indicate that the local HB network around the cis- conformers in the mixture is more cohesive in comparison with the corresponding local network around the trans- conformers.

**3.2 Hydrogen Bond Dynamics**

Apart from the static description of the hydrogen bonds formed in the cis/trans NMF liquid mixture, the HB dynamics were also investigated in the framework of the present study. According to the literature, the average dynamics for pairs i, j of hydrogen bonded atoms can be described using the following time correlation function (tcf) [56,57]:

$$C_{HB}(t) = \frac{\langle h_{ij}(0) \cdot h_{ij}(t) \rangle_{t^*}}{\langle h_{ij}(0)^2 \rangle} \tag{4}$$

The corresponding HB lifetime is defined as:

$$\tau_{HB} = \int_0^\infty C_{HB}(t) \cdot dt \qquad (5)$$

The variable $h_{ij}$ is defined such that $h_{ij}(t)=1$ when a hydrogen bond between specific atoms of the molecules i, j is formed at times 0 and t, and the same hydrogen bond has not been broken for a period longer than $t^*$, otherwise, $h_{ij}(t)=0$. Using this definition, the calculation of $C_{HB}(t)$ depends upon the selection of the parameter $t^*$. There are two limiting cases arising from this definition: a) When $t^* = 0$, which represents the so-called continuous definition, no bond ruptures and consequent reformations in the time interval [0, t] are observed. In this case, the calculated tcf is the continuous one $C_{HB}{}^C(t)$ and the corresponding lifetime is the continuous lifetime $\tau_{HB}{}^C$. b) When $t^* = \infty$, this represents the so-called intermittent definition. In this case the persistence probability at time t of a hydrogen bond created at t=0 is computed regardless of multiple ruptures and reformations of this bond during the time interval [0, t]. The calculated tcf is the intermittent one $C_{HB}{}^I(t)$ and the corresponding lifetime is the intermittent lifetime (or HB relaxation time) $\tau_{HB}{}^I$.

The calculated continuous and intermittent HB tcfs, $C_{HB}{}^C(t)$ and $C_{HB}{}^I(t)$, for all the investigated types of hydrogen bonds are presented in Figures 6 and 7. The continuous and intermittent lifetimes $\tau_{HB}{}^C$ and $\tau_{HB}{}^I$ of the investigated hydrogen bonds are also presented in Table 3. Note that for the calculation of the intermittent lifetimes, the HB tcfs were numerically integrated up to 20 ps and the rest of the integral was analytically calculated using a sum of three exponential decay functions (or a sum of two exponential decay functions in the case of the cis-cis $H_N\ldots O$ HB tcf) as fitting functions. predictions reveal that in general the continuous and intermittent HB tcfs of the $H_N\ldots O$ hydrogen bonds decay slower than the HB tcfs corresponding to $H_C\ldots O$ hydrogen bonds. The very fast continuous HB dynamics of the investigated $H_C\ldots O$ hydrogen bonds, as it can be clearly observed from the decay of $C_{HB}{}^C(t)$ tcfs and the calculated continuous HB lifetimes $\tau_{HB}{}^C$, are also reflected in the behavior of the corresponding calculated intermittent HB tcfs $C_{HB}{}^I(t)$. These tcfs, presented in the bottom panel of Figure 7, decay very fast in the 0-1 ps timescale, signifying that there is a strong interrelation between the continuous and the intermittent HB dynamics. They then exhibit a much slower decay, often observed in the case of intermittent HB dynamics. This behavior has also been observed in HB dynamics of ionic liquids [58] and it is a clear sign of the coupling between continuous and intermittent HB dynamics in strongly associated liquids, where the formation of hydrogen bonds is crucial for the determination of the system properties.

All these findings clearly indicate that the $H_N\cdots O$ hydrogen bonds persist for a longer timescale than the $H_C\cdots O$ ones. Interestingly the $H_N\cdots O$ hydrogen bond formed between cis- conformers exhibits the longest intermittent lifetime, indicating the strong local structural correlation among the cis molecules, which was also reflected by the intensity of the corresponding RDFs. Moreover, the intermittent HB lifetimes of the hydrogen bonds formed among cis- and trans- conformers are longer in comparison with the intermittent HB lifetimes of the hydrogen bonds formed among the trans- conformers. This behavior is more pronounced when the cis- conformers that form hydrogen bonds with trans- conformers act as HB acceptors. All the above findings clearly indicate that the local HB network around the cis- conformers is more cohesive and is maintained for longer timescales in comparison with the local HB network around the trans- conformers.

### 3.3 Reorientational Dynamics

The reorientational dynamics of the cis- and trans- conformers in the liquid mixture were investigated in terms of the well-known Legendre reorientational tcfs of specific adapted intramolecular vectors:

$$C_{lR}(t) = P_l \left\langle \vec{u}_i(0) \cdot \vec{u}_i(t) \right\rangle \qquad (6)$$

Here, $\vec{u}_i$ is a normalized vector associated to molecule i, and $P_l$ is a Legendre polynomial of order $l$. In the present study, the reorientational dynamics of intramolecular vectors associated to the formation of hydrogen bonds, namely the $N\text{-}H_N$, $C_C\text{-}H_C$ and $C_C\text{-}O$ vectors, were investigated. The calculated first, second and third order Legendre reorientational tcfs for all these vectors are presented in Figures 8-10. The corresponding reorientational correlation times have been calculated using the following equation:

$$\tau_{lR} = \int_0^\infty C_{lR}(t) \cdot dt \qquad (7)$$

The calculated correlation times, following the same methodology used to estimate the integrals in Eq. 7 as in the case of the HB lifetimes, are presented in Table 4. Figures 8-10 clearly show that the reorientational tcfs of the cis- conformers decay slower in comparison with the corresponding tcfs of the trans- conformers. This slower decay is reflected in the higher values obtained for the reorientational correlation times of the intramolecular vectors of the cis- conformers. The slower reorientation of the cis- conformers can be explained in terms of the more cohesive local HB network around them, which is also in agreement with previous findings reporting that the translational motions of the cis- conformers are also slower in the mixture [30].

The calculated second order Legendre reorientational correlation time $\tau_{2R}$ for the $C_C$-$H_C$ vector of trans-NMF (the dominating component in the liquid mixture) is 6.5 ps, in reasonable agreement with the experimental values of 5.9 and 4.7 ps reported in previous NMR studies by Chapman et al [59] and Seipelt and Zeidler [60], respectively. All these findings clearly indicate that our BOMD simulation, apart from the local structural properties of the mixture, also provides reliable descriptions of its dynamic properties.

From the calculated values of the reorientational correlation times, presented in Table 4, it can also be observed that the values obtained for the N-$H_N$, $C_C$-$H_C$ and $C_C$-O vectors in the case of the cis- conformers are similar. On the other hand, in the case of trans- conformers the reorientational dynamics of the $C_C$-$H_C$ intramolecular vector are much faster, probably due to the fact the dynamics of the $H_C$…O hydrogen bonds formed among trans- conformers molecule and their closest trans- or cis- conformer neighbors are also much faster and therefore control the reorientation of the $C_C$-$H_C$ intramolecular vector. The strong interrelation between single reorientational and HB dynamics is reflected on the similarity of the calculated first-order Legendre reorientational correlation times of the N-$H_N$ vector and the corresponding intermittent HB lifetimes of the $H_N$…O hydrogen bonds formed between a trans- conformer and its closest trans- or cis- conformer neighbor, presented in Tables 3 and 4.

Deviations of the calculated second- and third-order Legendre reorientational correlation times for both the cis- and trans-NMF molecules, particularly in the case of the third-order ones, from the trends expected when using the Debye diffusion model of reorientational relaxation [61] ($\tau_{1R} = 3 \cdot \tau_{2R}$, $\tau_{1R} = 6 \cdot \tau_{3R}$) have also been observed. These deviations are more pronounced for the $C_C$-$H_C$ and $C_C$-O vectors, signifying that the reorientation dynamics of these particular vectors are less diffusive in comparison with the dynamics of the N-$H_N$ intramolecular vector. This can be attributed to the much faster continuous and intermittent HB dynamics of the $H_C$...O hydrogen bonds. These deviations from the Debye diffusion model of molecular reorientation, as also pointed out for HB liquids such as water [62], provide additional indications of the strong effect of HB interactions on the single molecule reorientational dynamics in associated liquids.

**3.4 Intermolecular-Intramolecular Vibrational Dynamics**

The translational dynamics of the cis- and trans- conformers was further investigated by means of the atomic velocity correlation functions $C_v^i(t)$ of their $H_N$, $H_C$ and oxygen atoms and their corresponding spectral densities $S_v^i(\omega)$, calculated by performing a Fourier transform of the velocity tcfs:

$$C_v^i(t) = \frac{\left\langle \vec{v}_i(0) \cdot \vec{v}_i(t) \right\rangle}{\left\langle \vec{v}_i(0)^2 \right\rangle} \qquad i = H_N, H_C, O \qquad (8)$$

$$S_v^i(\omega) = \int_0^\infty \cos(\omega \cdot t) \cdot C_v^i(t) \cdot dt \qquad (9)$$

The spectral densities $S_v^i(\omega)$ were calculated by numerical integration using a Boole's rule and applying a Hanning window to eliminate the truncation effects in the calculation of the Fourier transform of the calculated tcfs [55]. The calculated spectral densities for the cis- and trans-NMF are presented in Figure 11. In the top panel of Figure 11 the low-frequency spectrum, which provides information about local cage effects and intermolecular vibrations due to HB interactions, is presented. All the calculated spectral densities exhibit the typical peaks due to cage effects in liquids [63] at very low frequencies in the range 40-60 cm$^{-1}$. However, the more compact local HB network around the cis- conformer is reflected in the peaks observed in the spectral densities of the $H_N$ and $H_C$ velocity tcfs located at 146 and 151 cm$^{-1}$, respectively. These peaks are blue-shifted in comparison with the peaks observed in the spectral densities of the $H_N$ and $H_C$ velocity tcfs of the trans- conformers located at 88 and 96 cm$^{-1}$, respectively. The observation of this blueshift in the low-frequency part of the spectrum is indicative of a more cohesive HB network around the cis- conformers, as it is also well known for water [54, 63] and other hydrogen-bonded liquids [55]. This blueshift is not observed in the case of the spectral densities of the oxygen velocity tcfs, signifying that the motions of the hydrogen atoms are more sensitive to the local environment around the two conformers in the liquid mixture.

On the other hand, the high-frequency part of the calculated spectral densities, presented in the bottom panel of Figure 11, depicts the intramolecular N-$H_N$ and $C_C$-$H_C$ bond stretching vibrations, which according to the literature are also sensitive to the local HB network around the molecules in a liquid. Our previous Car-Parrinello MD studies [63] on liquid ambient water revealed that when the number of hydrogen bonds formed by a water molecule in the liquid increases, the peak corresponding to the O-H bond stretching vibration is red-shifted. In the present study we reveal that the peak corresponding to the N-$H_N$ bond stretching vibration in the case of the trans- conformers is located at 3337 cm$^{-1}$ and in the case of the cis- conformers it is redshifted at 3299 cm$^{-1}$. Similarly, the peak corresponding to the $C_C$-$H_C$ bond stretching vibration in the case of the trans- conformers is located at 2933 cm$^{-1}$ and in the case of the cis- conformers it is redshifted at 2925 cm$^{-1}$. This red-shift in the intramolecular N-$H_N$ and $C_C$-$H_C$ bond stretching vibrations observed for the cis- conformers is in line with our previous findings for water and provides further evidence for the existence of a more cohesive local HB network around the cis- conformers in the liquid mixture.

**3.5 Liquid State Dipole Moment**

Since it is commonly known that polarization effects in polar molecular liquids are very important, in the framework of the present study we calculated the dipole moment of the cis- and trans- conformers in the liquid mixture. The molecular dipole moments were estimated by calculating the maximally localized Wannier functions [64,65]. These functions were used to partition the total charge density of the system into single-molecule contributions. The dipole moment of each molecule was calculated from the ionic and the Wannier function center positions [66-68], using the TRAVIS software [69]. The probability density distributions for the magnitudes of the dipole moment vectors of the cis- and trans- conformers are presented in Figure 12. The average value of the molecular dipole moment of trans-NMF is 6.21 D (with a standard deviation of 0.62 D), whereas the corresponding average value for cis-NMF is 6.22 D (with a standard deviation of 0.53 D). The experimental gas-phase value for trans-NMF is 3.78 D, whereas previous *ab initio* studies estimated the gas-phase dipole moments of cis- and trans-NMF to be 4.14 and 4.29 D [70], respectively. These findings emphasize the importance of polarization phenomena in liquid NMF. Interestingly, previous *ab initio* MD simulation studies of liquid trans-NMA [71], which is very similar to NMF, estimated a liquid-state dipole moment of 6.0 D, whereas the experimental and calculated dipole moment values were 3.73 and 3.74 D, respectively. These results further validate our present findings, indicating that induction interactions play a very important role in polar liquid amide solvents.

**4. Conclusions**

Born-Oppenheimer DFT-based MD simulations were employed to investigate the local structure, HB interactions and related dynamics in the cis/trans NMF liquid mixture. The calculated RDFs and neutron-weighted structure factor are in good agreement with previous experimental neutron diffraction and reverse modelling studies and provide a systematic description of the local intermolecular structure around the two conformers in the liquid. The average total number of hydrogen bonds formed per cis-NMF molecules with its closest cis- and trans-NMF neighbors is 2.45, whereas the corresponding average total number per trans-NMF molecules is predicted to be 2.2. Moreover, the investigation of the dynamics of the hydrogen bonds formed in the liquid revealed that the local HB network around the cis- conformers it is maintained for longer timescales in comparison with the local HB network around the trans- conformers. The more cohesive local HB network around the cis-NMF molecules causes a corresponding slowing-down of the reorientational dynamics of these molecules. The observed local structural effects are also reflected in the spectral densities of the atomic velocity tcfs of the $H_N$ and $H_C$ hydrogen atoms of cis- conformers. More specifically, we revealed that the compact HB network around the cis- conformers causes a corresponding blue-shift of the low-frequency intermolecular vibrations due to the formation of hydrogen bonds and a red-shift of the intramolecular N-$H_N$ and $C_C$-$H_C$ bond stretching vibrations, which is in line with our previous works regarding the effect of the local HB network in liquid water on its intermolecular and intramolecular vibrational motions. Finally, the present work revealed that the dipole moments of the cis- and trans- conformers are significantly higher in the liquid state, compared with their gas-phase values. The latter finding highlights the importance of polarization effects in this particular polar liquid solvent.

**Acknowledgements**

This work was supported by computational time granted from the National Infrastructures for Research and Technology S.A. (GRNET S.A.) in the National HPC facility - ARIS - under project ID pr002018 (Project's acronym: FPMDCISTRANS). Fruitful discussions with Professor Alan K. Soper (Rutherford Appleton Laboratory, UK) regarding his previous experimental and reverse modelling studies of liquid NMF are also gratefully acknowledged.

# TABLES

**Table1:** Calculated average number of hydrogen bonds formed per trans- conformer in the liquid, for each different case where the trans- conformer acts as a HB donor or acceptor.

| Molecular Pairs | trans-trans | | | |
|---|---|---|---|---|
| H. Bonds Formed | $H_N \ldots O$ | $O \ldots H_N$ | $H_C \ldots O$ | $O \ldots H_C$ |
| $< n_{HB} >$ (per trans molecule) | 0.80 | 0.80 | 0.20 | 0.20 |
| Molecular Pairs | trans-cis | | | |
| H. Bonds Formed | $H_N \ldots O$ | $O \ldots H_N$ | $H_C \ldots O$ | $O \ldots H_C$ |
| $< n_{HB} >$ (per trans molecule) | 0.08 | 0.07 | 0.02 | 0.03 |

**Table2:** Calculated average number of hydrogen bonds formed per cis- conformer in the liquid, for each different case where the cis- conformer acts as a HB donor or acceptor.

| Molecular Pairs | cis-cis | | | |
|---|---|---|---|---|
| H. Bonds Formed | $H_N \ldots O$ | $O \ldots H_N$ | $H_C \ldots O$ | $O \ldots H_C$ |
| $< n_{HB} >$ (per cis molecule) | 0.25 | 0.25 | 0.02 | 0.02 |
| Molecular Pairs | cis-trans | | | |
| H. Bonds Formed | $H_N \ldots O$ | $O \ldots H_N$ | $H_C \ldots O$ | $O \ldots H_C$ |
| $< n_{HB} >$ (per cis molecule) | 0.65 | 0.79 | 0.27 | 0.20 |

**Table 3:** Calculated values of the continuous and intermittent lifetimes, $\tau_{HB}^{C}$ and $\tau_{HB}^{I}$, of the investigated hydrogen bond types.

| H. Bond Type | $\tau_{HB}^{C}$ (ps) | $\tau_{HB}^{I}$ (ps) |
|---|---|---|
| $H_N\ldots O$ (trans-trans) | 0.83 | 31.7 |
| $H_N\ldots O$ (cis-trans) | 0.64 | 34.5 |
| $H_N\ldots O$ (trans-cis) | 0.94 | 41.1 |
| $H_N\ldots O$ (cis-cis) | 0.72 | 67.9 |
| $H_C\ldots O$ (trans-trans) | 0.11 | 5.2 |
| $H_C\ldots O$ (cis-trans) | 0.13 | 11.6 |
| $H_C\ldots O$ (trans-cis) | 0.09 | 22.8 |

**Table 4:** Calculated first-, second- and third-order Legendre reorientational correlation times for the N-$H_N$, $C_C$-$H_C$ and $C_C$-O intramolecular vectors of the trans- and cis-conformers.

| trans-NMF | | | |
|---|---|---|---|
| Vector | $\tau_{1R}$ (ps) | $\tau_{2R}$ (ps) | $\tau_{3R}$ (ps) |
| N-$H_N$ | 38.2 | 12.7 | 5.7 |
| $C_C$-$H_C$ | 18.1 | 6.5 | 3.2 |
| $C_C$-O | 42.1 | 13.1 | 5.8 |
| cis-NMF | | | |
| Vector | $\tau_{1R}$ (ps) | $\tau_{2R}$ (ps) | $\tau_{3R}$ (ps) |
| N-$H_N$ | 51.6 | 17.8 | 7.8 |
| $C_C$-$H_C$ | 55.0 | 17.2 | 7.4 |
| $C_C$-O | 58.5 | 16.5 | 7.3 |

**FIGURES**

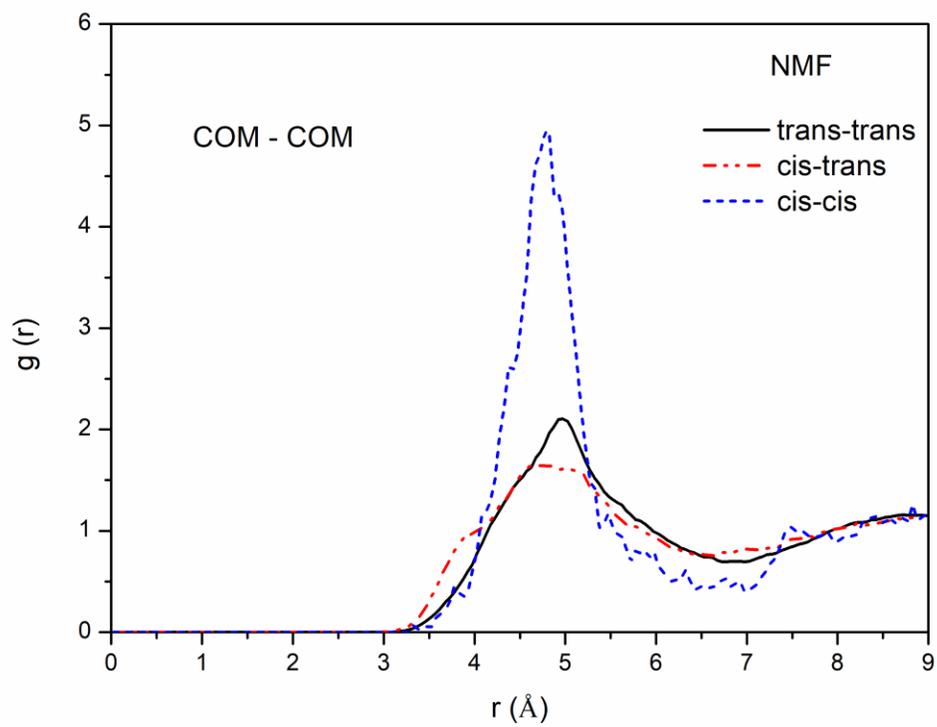

**Figure 1**

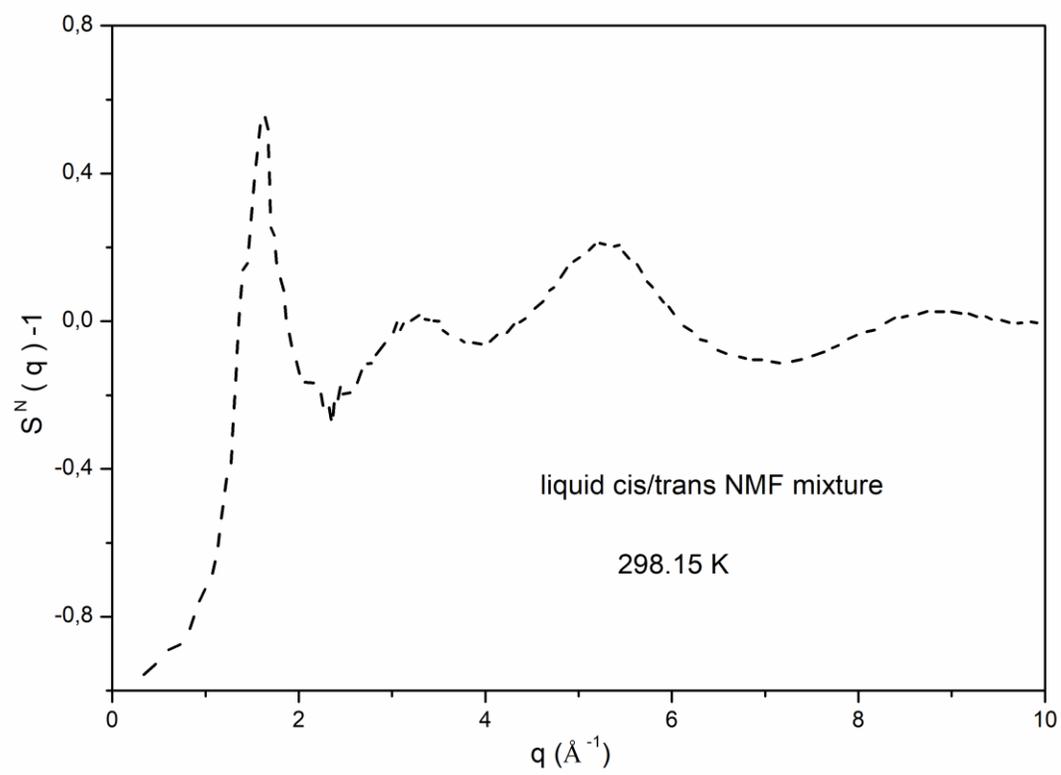

**Figure 2**

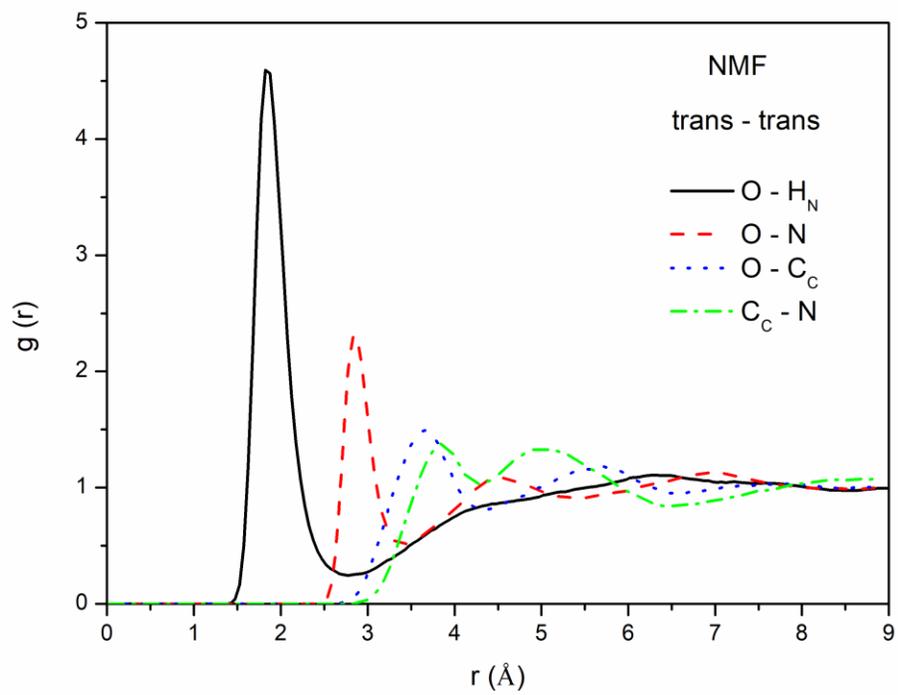
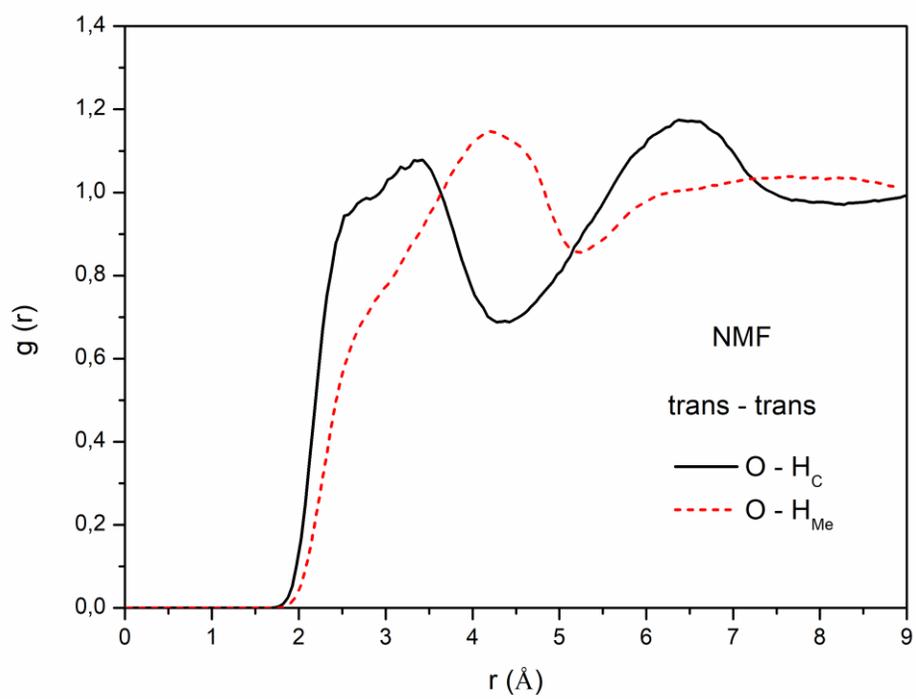

**Figure 3**

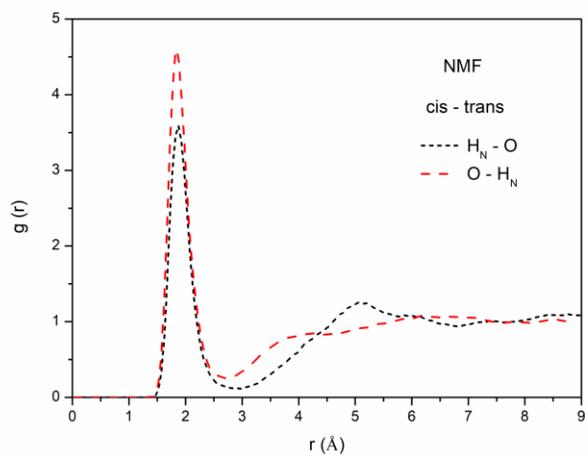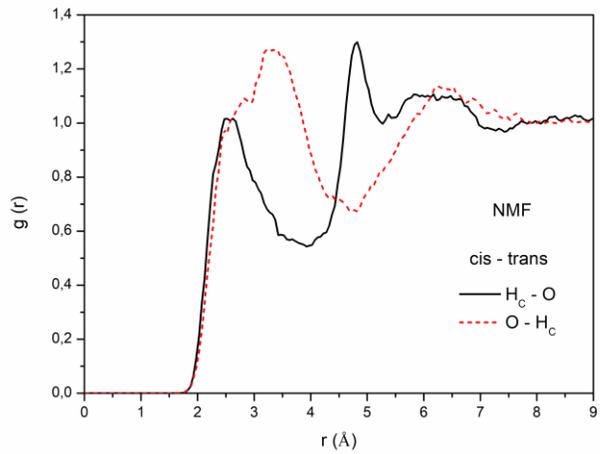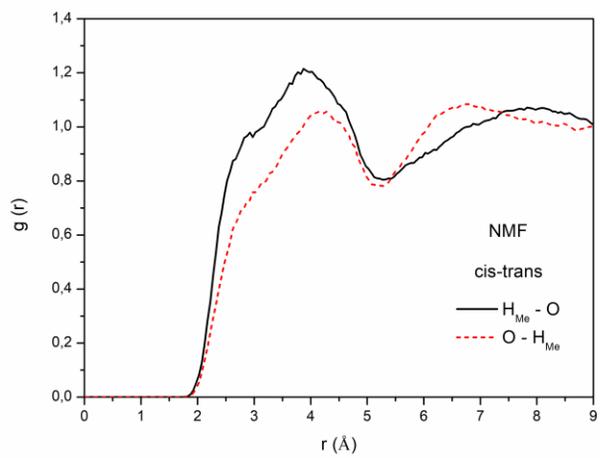

**Figure 4**

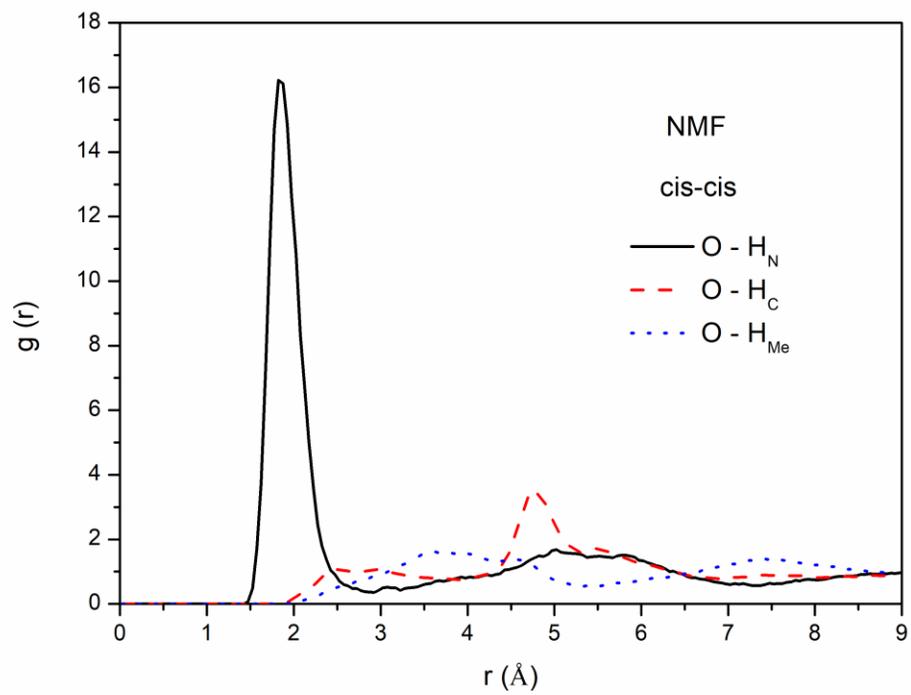

**Figure 5**

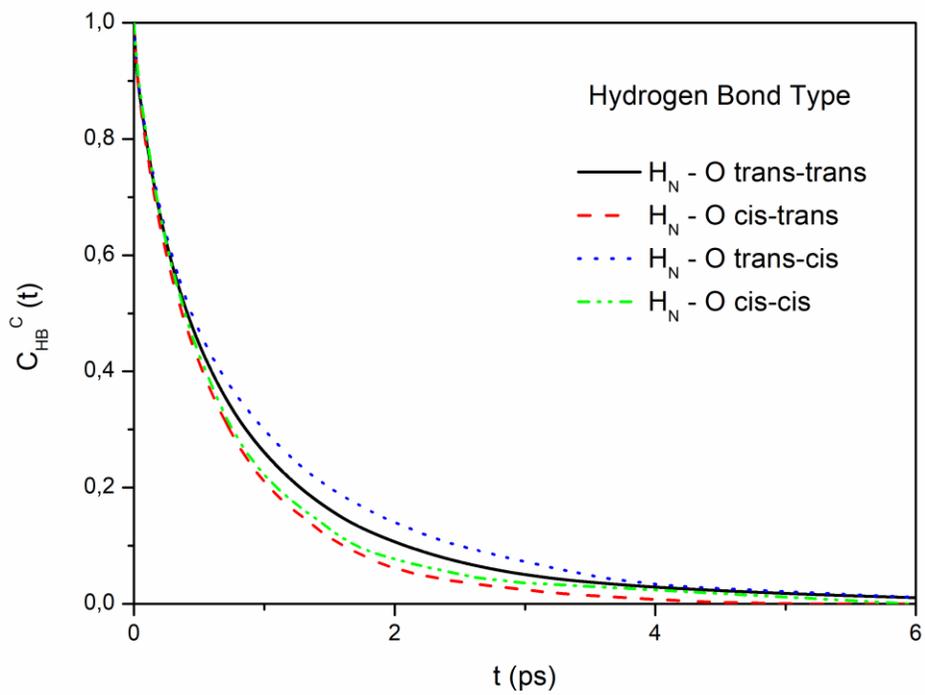

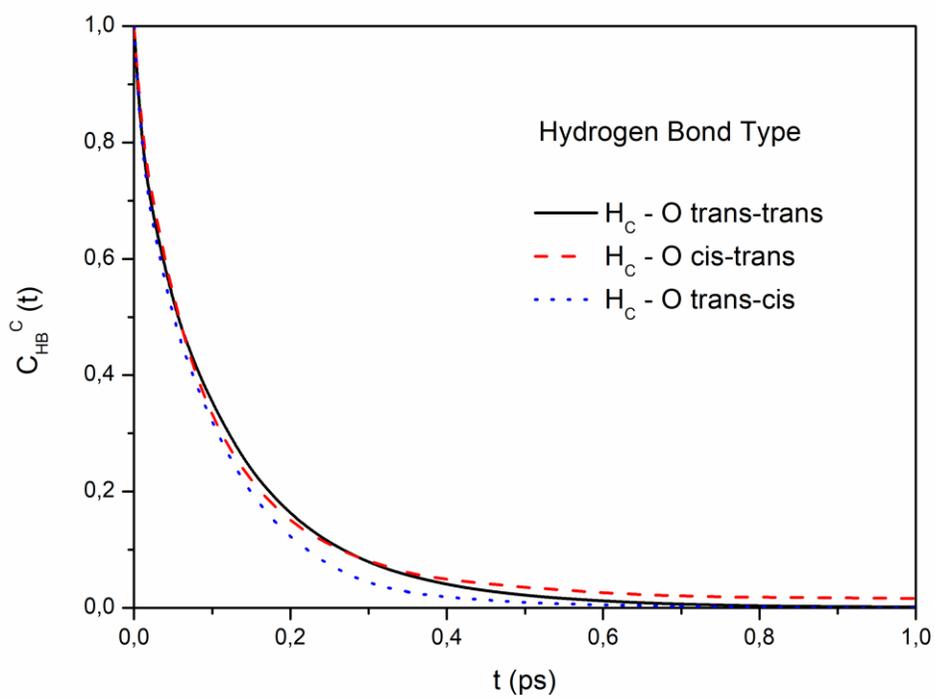

**Figure 6**

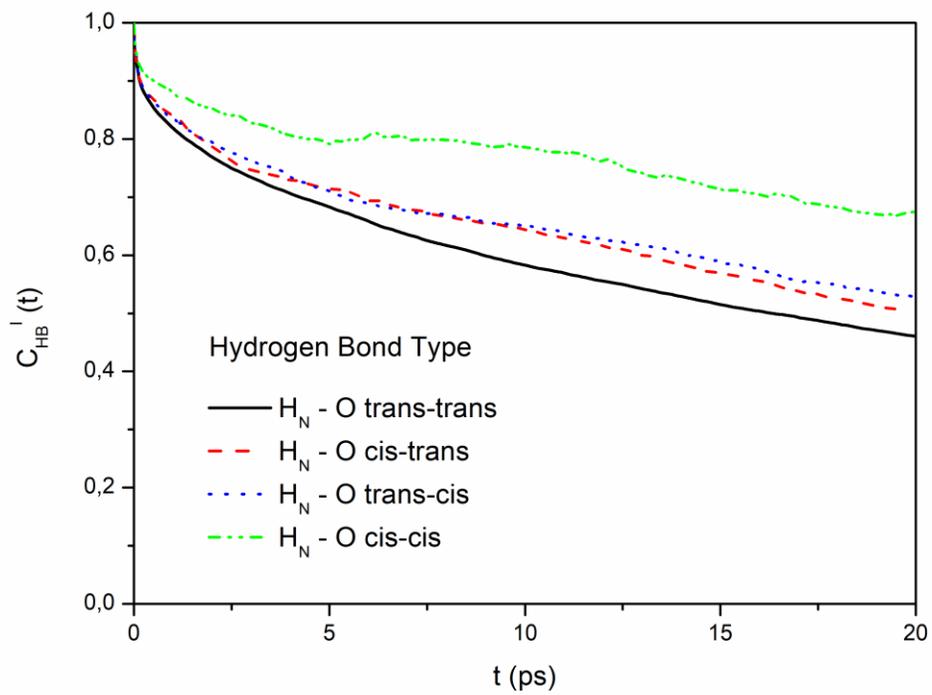

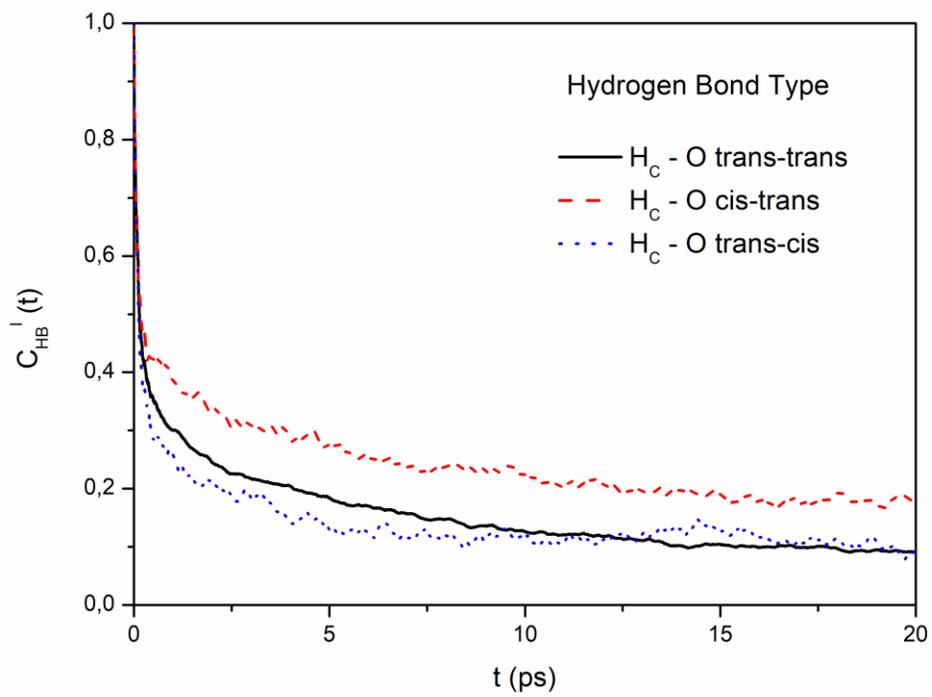

**Figure 7**

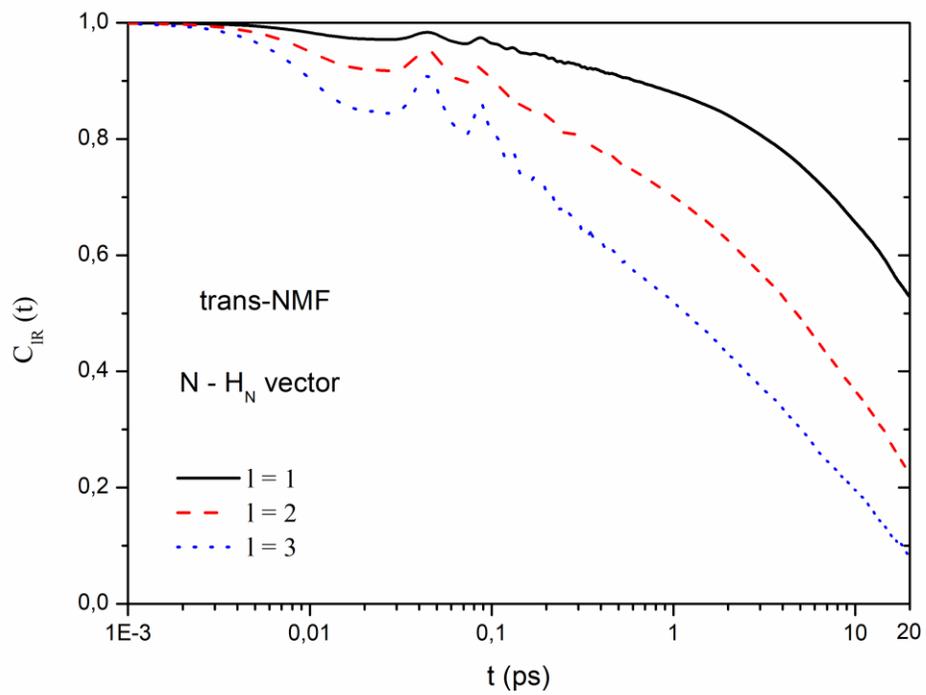

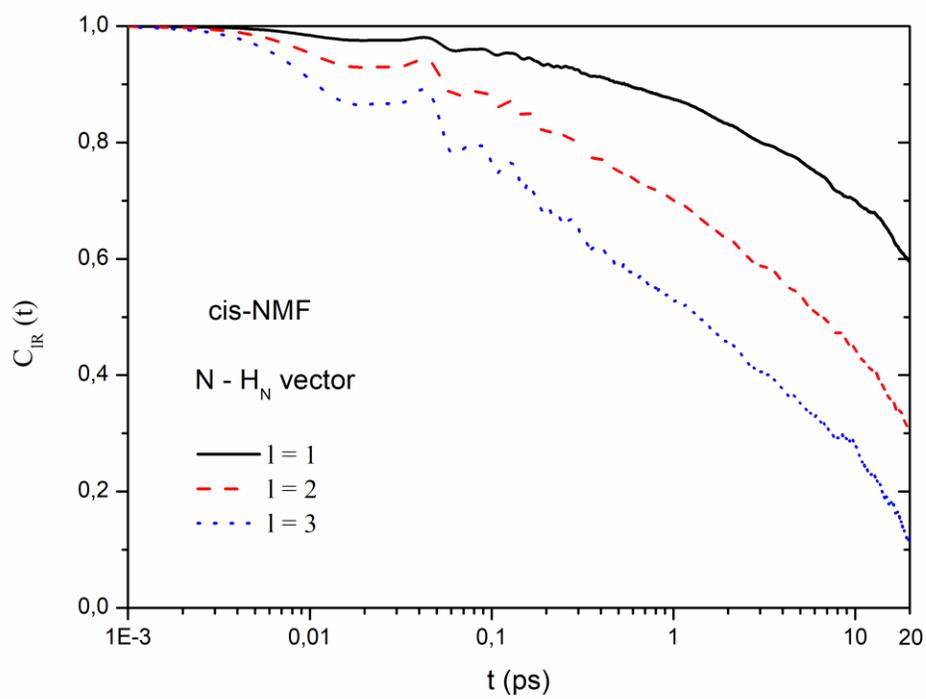

**Figure 8**

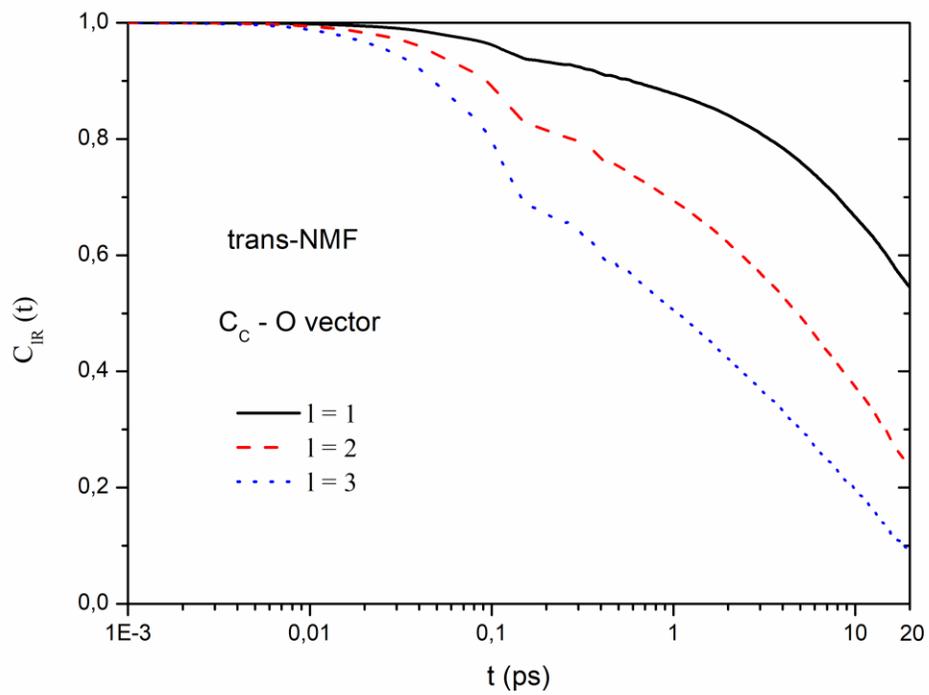
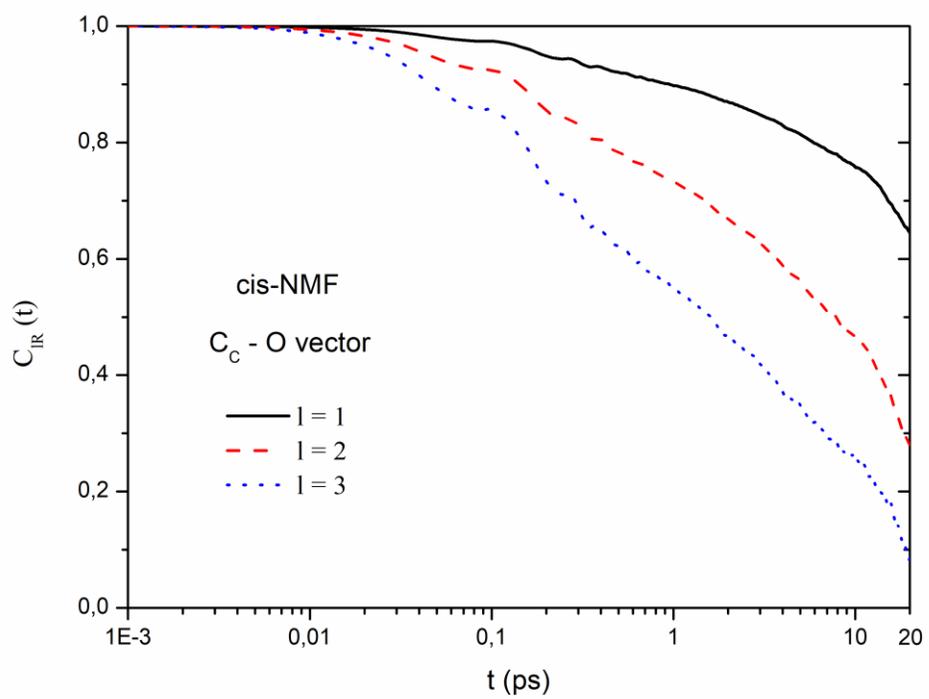

**Figure 9**

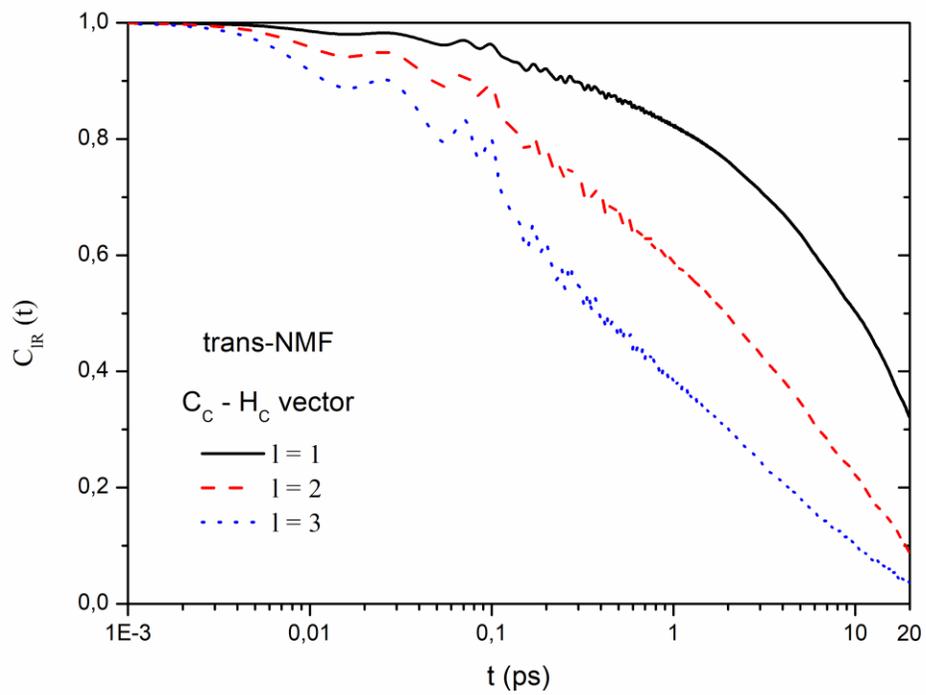

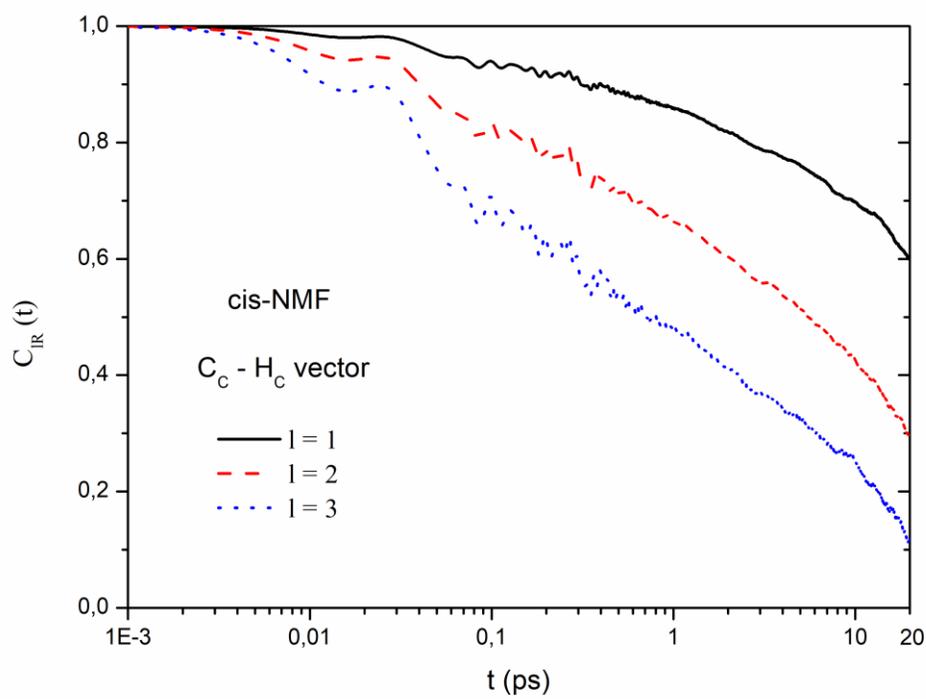

**Figure 10**

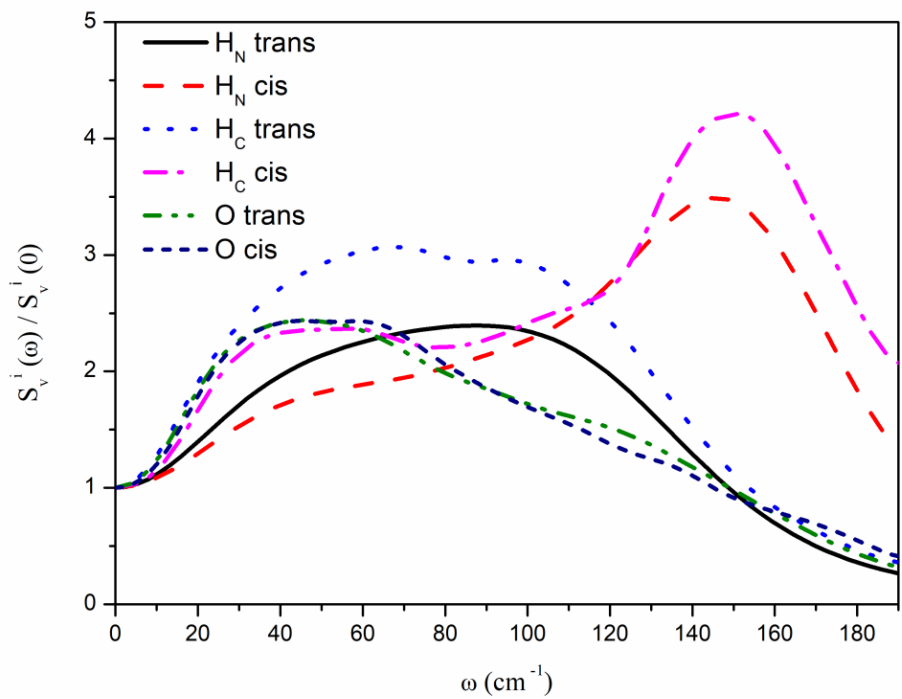

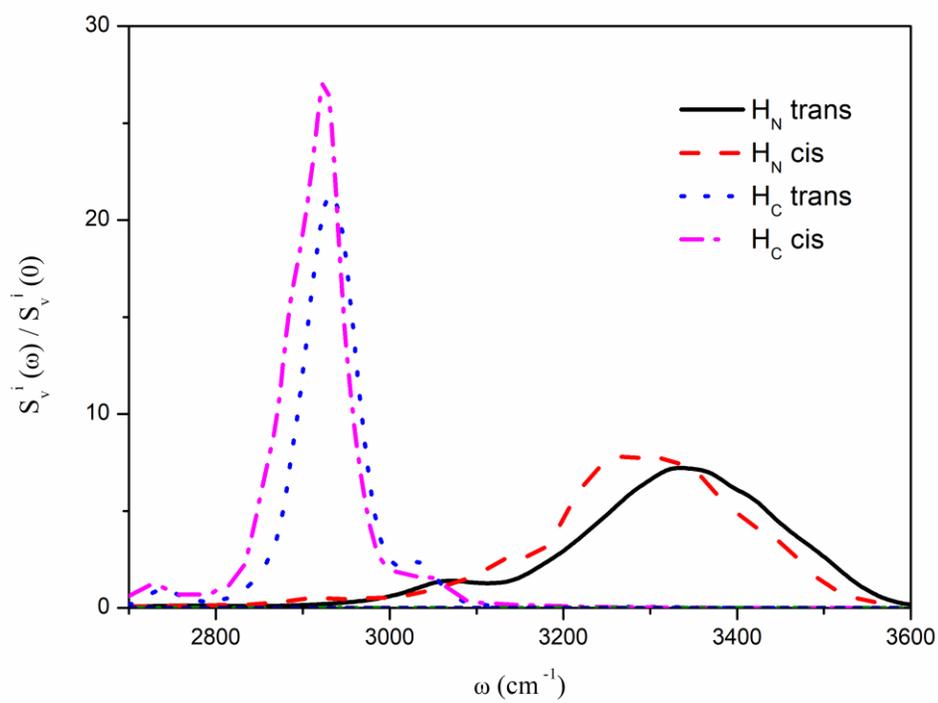

**Figure 11**

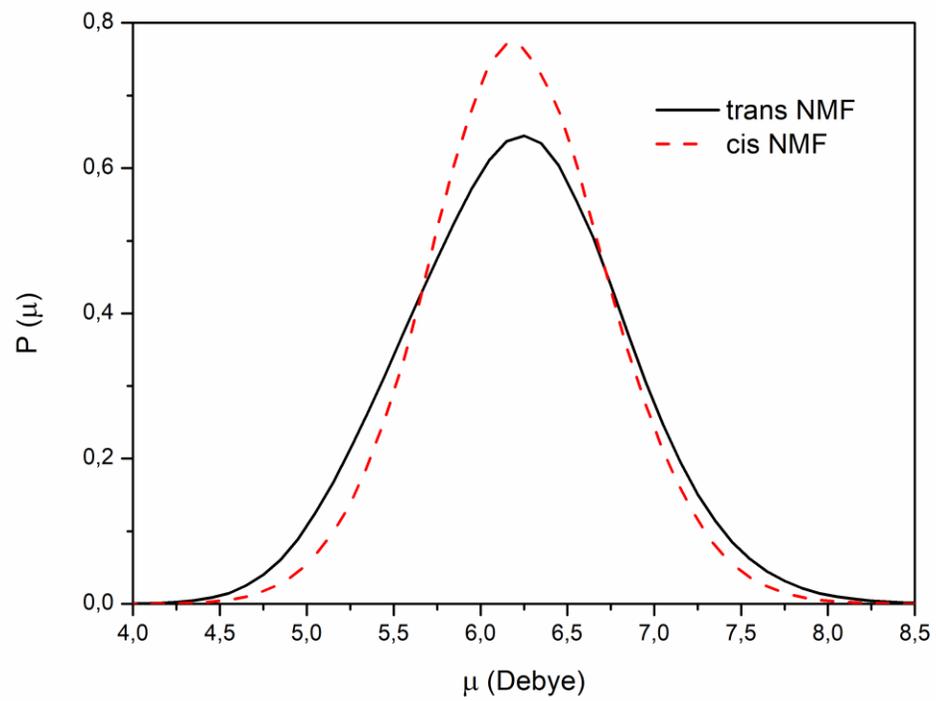

**Figure 12**

# FIGURE CAPTIONS

**Figure 1:** Calculated COM-COM RDF for the trans-trans, cis-trans and cis-cis NMF pairs.

**Figure 2:** Calculated neutron-weighted structure factor, $S^N(q)-1$.

**Figure 3:** Calculated atom-atom RDF for the trans-trans NMF pairs.

**Figure 4:** Calculated atom-atom RDF for the cis-trans NMF pairs.

**Figure 5:** Calculated atom-atom RDF for the cis-cis NMF pairs.

**Figure 6:** Calculated continuous HB tcfs, $C_{HB}^{C}(t)$, for the investigated hydrogen bonds.

**Figure 7:** Calculated intermittent HB tcfs, $C_{HB}^{I}(t)$, for the investigated hydrogen bonds.

**Figure 8:** Calculated first-, second- and third-order Legendre reorientational tcfs for the N-$H_N$ intramolecular vector of the trans- and cis- conformers.

**Figure 9:** Calculated first-, second- and third-order Legendre reorientational tcfs for the $C_C$-O intramolecular vector of the trans- and cis- conformers.

**Figure 10:** Calculated first-, second- and third-order Legendre reorientational tcfs for the $C_C$-$H_C$ intramolecular vector of the trans- and cis- conformers.

**Figure 11:** Low- and high-frequency regions of the calculated spectral densities of the atomic velocity tcfs for the $H_N$, $H_C$ and O atoms of the trans- and cis- conformers.

**Figure 12:** Calculated probability density distributions for the magnitudes of the dipole moment vectors of the trans- and cis- conformers.